# Achieving a uniform static magnetic field over 50T in a large free space

Fei Sun[1,2] and Sailing He[1,2]*

1 Department of Electromagnetic Engineering, School of Electrical Engineering, Royal Institute of Technology (KTH), S-100 44 Stockholm, Sweden

2 Centre for Optical and Electromagnetic Research, Zhejiang Provincial Key Laboratory for Sensing Technologies, JORCEP, East Building #5,Zijingang Campus, Zhejiang University, Hangzhou 310058, China

* Corresponding author: sailing@jorcep.org

Abstract:

  We propose a compact passive device as a super-concentrator to obtain an extremely high uniform static magnetic field over 50T in a large two-dimensional free space in the presence of a uniform weak background magnetic field. Our design is based on transformation optics and metamaterials for static magnetic fields. Finite element method (FEM) is utilized to verify the performance of the proposed device.

1. Introduction and background

In recent years, many novel optical devices have been designed and experimentally demonstrated with transformation optics (TO) [1-4]. TO is based on the invariance form of Maxwell's equations, which can give a one-to-one mapping between the two spaces (a reference space and a physical space). By choosing appropiate transformation functions and reference space, one can design some optical devices with peculiar functionalities, such as invisible cloaks [5, 6], concentrators [7], field rotators [8], polarization controllers [9], optical black holes [10, 11], illusion optical devices [12], etc.

  More and more attention has recently focused on controlling static fields by TO. Pendry et al have used TO to design plasmonic devices in quasi-static condition [13]. A DC electric cloak [14] and a DC electric concentrator [15] have been proposed and experimentally realized very recently. In the development of DC magnetic metamaterials [16, 17], many DC magnetic devices based on TO (e.g., DC magnetic cloaking [18, 19] and DC magnetic energy concentrators [20]) have also been studied in recent years.

  In fact, the static or quasi-static magnetic fields play an essential role in many applications, including metal detection [19], magnetic sensing [21], and wireless energy transmission [22]. Concentrating a DC magnetic field to obtain a higher magnetic field in free space has important applications for increasing the sensitivity of a magnetic sensor [23, 24], improving the performance of transcranial magnetic stimulation [25, 26], and improving the spatial resolution of magnetic resonance imaging (MRI) [27], etc. High magnetic fields have been used in many areas, including physics, biology, materials science, and engineering. Very high magnetic fields are usually generated by resistive and/or superconducting magnet systems [28, 29]. However, energizing such magnets requires an extremely high electrical power. So far the strongest artificial DC magnetic field we human being can create is no more than 45T (created at the National High Magnetic Field Laboratory in the United States, which consumes an electrical power of about 30 MW [30]).

  In this paper, we will summarize how to use TO to enhance static magnetic fields and design a super-concentrator, which is a passive compact device and can realize a DC magnetic field larger than 50T inside it (free space region) in the presence of a uniform weak background magnetic field. Note that this is different from the enhancement of harmonic electromagnetic field. If one uses the device proposed in [20] to obtain a large DC magnetic field over 50T in 2D free space, the ratio of outer radius to inner radius of the device in [20] should be larger than 50, which means we have to sacrifice space greatly to obtain a high magnetic field. Based on folded transformation, our super-concentrator can achieve an extremely high magnetic field in a large free space with a very small ratio of outer radius to inner radius.

We will first review TO for static magnetic fields. Then we will summarize different methods for enhancing static magnetic fields by using TO, and finally propose a super-concentrator for static magnetic fields. Finite element method is used to verify our ideas, and comparison with other methods is also given.

2. Transformation optics for static magnetic fields

We will first summarize TO for static magnetic fields. As a static field is the limiting case when $\omega \to 0$ in Maxwell's equations, TO for the static field can directly be derived from the general TO formula by setting $\omega \to 0$. In this case, the electric field and magnetic field are decoupled. We can separately design permittivity and permeability by coordinate transformation to control the static electric field and magnetic field. As Maxwell's equations are invariant under the coordinate transformation, divergence and curl equations for the static magnetic field are also invariant under the coordinate transformation. The equations for the static magnetic field in a source free space can be written as:

$$\overline{\nabla} \times \overline{H} = 0, \qquad \overline{\nabla} \cdot \overline{\overline{\mu}} \overline{H} = 0. \tag{1}$$

The coordinate transformation can be expressed by using Jacobian transformation matrix $M$. If we want to keep the form of Eq. (1) invariant, the field and material should be transformed accordingly by:

$$\overline{\overline{\mu}}' = M \overline{\overline{\mu}} M^T / \det(M), \overline{H}' = (M^T)^{-1} \overline{H}. \tag{2}$$

Note here that quantities in the physical space and reference space are indicated with and without superscript, respectively.

3. Enhance a static magnetic field with TO

We can give various ways to enhance a DC magnetic field by using TO. For example, we can simply compress a big volume to a small one while keeping the outside an identical transformation to enhance the field (see Fig. 1 (a) and (b)), as proposed in [7, 20]. Using this configuration, if we want to obtain an extremely large DC magnetic field, we should drastically compress the space. This means that if we want to obtain a very high DC magnetic field in a small region, the size of the device based on this method should be very large (see Fig. 2 (a)). We can also use TO based on mimicking celestial mechanics [10] or other non-Euclidean spaces [6] to design a magnetic black hole or a magnetic harvester for concentrating the DC magnetic field. However, these methods can only concentrate the DC magnetic field inside a magnetic material, but not in free space (see Fig. 2(b)). Inspired by superscatterers designed by TO for enlarging the scattering cross section [31, 32], here we propose to use a space-folded transformation to concentrate the DC magnetic field. By using this transformation, we can obtain a uniform DC magnetic field larger than 50T in 2D free space within a very compact passive device (see Fig. 2(c) and (d) below).

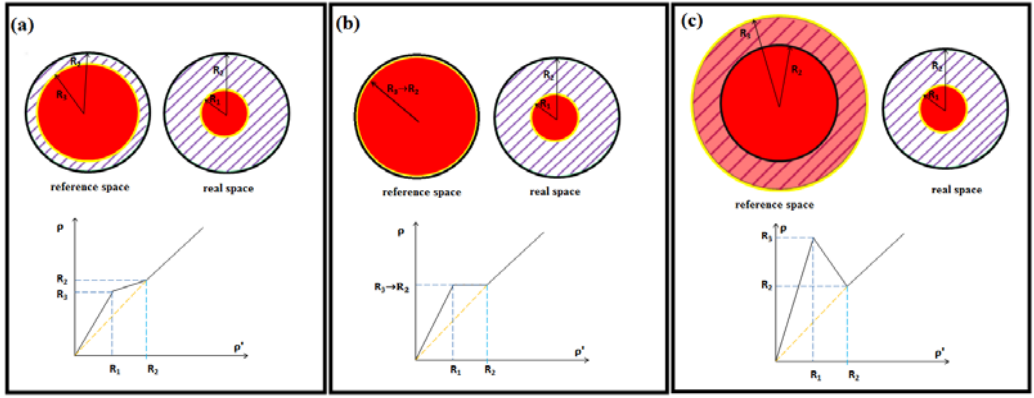

Fig. 1. The linear transformation between the reference space and the physical space: a traditional concentrator (a), a magnetic energy harvesting device in [20] (b), and the present super-concentrator (c).

We consider a cylindrical shell (infinitely long in the z direction) with inner radius $R_1$ and outer radius $R_2$. The proposed transformation can be written as

$$\rho' = \begin{cases} \dfrac{R_1}{R_3}\rho, & \rho' \in [0, R_1) \\ \dfrac{R_2 - R_1}{R_2 - R_3}\rho + R_2 \dfrac{R_1 - R_3}{R_2 - R_3}, & \rho' \in [R_1, R_2]; \quad \theta' = \theta; \quad z' = z. \\ \rho, & \rho' \in (R_2, \infty) \end{cases} \quad (3)$$

The Jacobian transformation matrix can be calculated as

$$M = \begin{cases} diag\left(\dfrac{R_1}{R_3}, \dfrac{R_1}{R_3}, 1\right), & \rho' \in [0, R_1) \\ diag\left(\dfrac{R_2 - R_1}{R_2 - R_3}, \dfrac{R_2 - R_1}{R_2 - R_3} \dfrac{\rho'}{\rho' - R_2 \dfrac{R_1 - R_3}{R_2 - R_3}}, 1\right), & \rho' \in [R_1, R_2]. \\ 1, & \rho' \in (R_2, \infty) \end{cases} \quad (4)$$

The transformation medium can be calculated with Eq. (2) as

$$\mu' = \begin{cases} diag\left(1, 1, \left(\dfrac{R_3}{R_1}\right)^2\right), & \rho' \in [0, R_1) \\ diag\left(\dfrac{\rho' - R_2 \dfrac{R_1 - R_3}{R_2 - R_3}}{\rho'}, \dfrac{\rho'}{\rho' - R_2 \dfrac{R_1 - R_3}{R_2 - R_3}}, \left(\dfrac{R_2 - R_1}{R_2 - R_3}\right)^2 \dfrac{\rho' - R_2 \dfrac{R_1 - R_3}{R_2 - R_3}}{\rho'}\right), & \rho' \in [R_1, R_2]. \\ 1, & \rho' \in (R_2, \infty) \end{cases} \quad (5)$$

For a two-dimensional (2D) case (here the magnetic field is in the plane), only $\mu_r$ and $\mu_\theta$ effect the magnetic field. We can rewrite Eq. (5) as

$$\mu' = \begin{cases} 1, \rho' \in [0, R_1) \\ diag(\dfrac{\rho' - R_2 \dfrac{R_1 - R_3}{R_2 - R_3}}{\rho'}, \dfrac{\rho'}{\rho' - R_2 \dfrac{R_1 - R_3}{R_2 - R_3}}), \rho' \in [R_1, R_2]. \\ 1, \rho' \in (R_2, \infty) \end{cases} \quad (6)$$

As we can see the device is filled in region $\rho' \in [R_1, R_2]$. In the 2D case it is air both inside and outside of the device. For a traditional concentrator, it requires $R_1 < R_3 < R_2$ (see Fig. 1(a)). For the magnetic energy harvesting device recently proposed in [20], it still satisfies $R_1 < R_3 < R_2$ but with the only difference being that $R_3$ infinitely approaches $R_2$ (see Fig. 1(b)). This is also the reason why they obtain $\mu_\theta \to 0$ and $\mu_\rho \to \infty$. For our proposed super-concentrator, we have $R_1 < R_2 < R_3$ (see Fig. 1(c)). In this case, the degree of concentration is not limited by the size of device. However, we have to introduce negative anisotropic permeability for the device.

We use the finite element method (FEM) to verify our design. As shown in Fig. 2(c) and (d), if we apply a uniform background static magnetic field whose magnetic flux density is 1T (as a numerical example), the static magnetic field inside our device (free space) can be larger than 50T, which cannot be achieved by any traditional method [29, 30]. The material parameters of our device are shown in Fig. 3. For comparison, we also use FEM to show the enhancement of the static magnetic field achieved with other TO-based methods (see Fig. 2(a) and (b)). As we have mentioned before, if we use the magnetic harvesting device designed in [20], we have to use a device whose geometrical size is extremely large to obtain a high magnetic field in a small region (see Fig. 2(a)). We can also design a magnetic black hole to enhance the static magnetic field (see Fig. 2(b)). However, this method (or other similar methods) cannot achieve a uniform static magnetic field in free space.

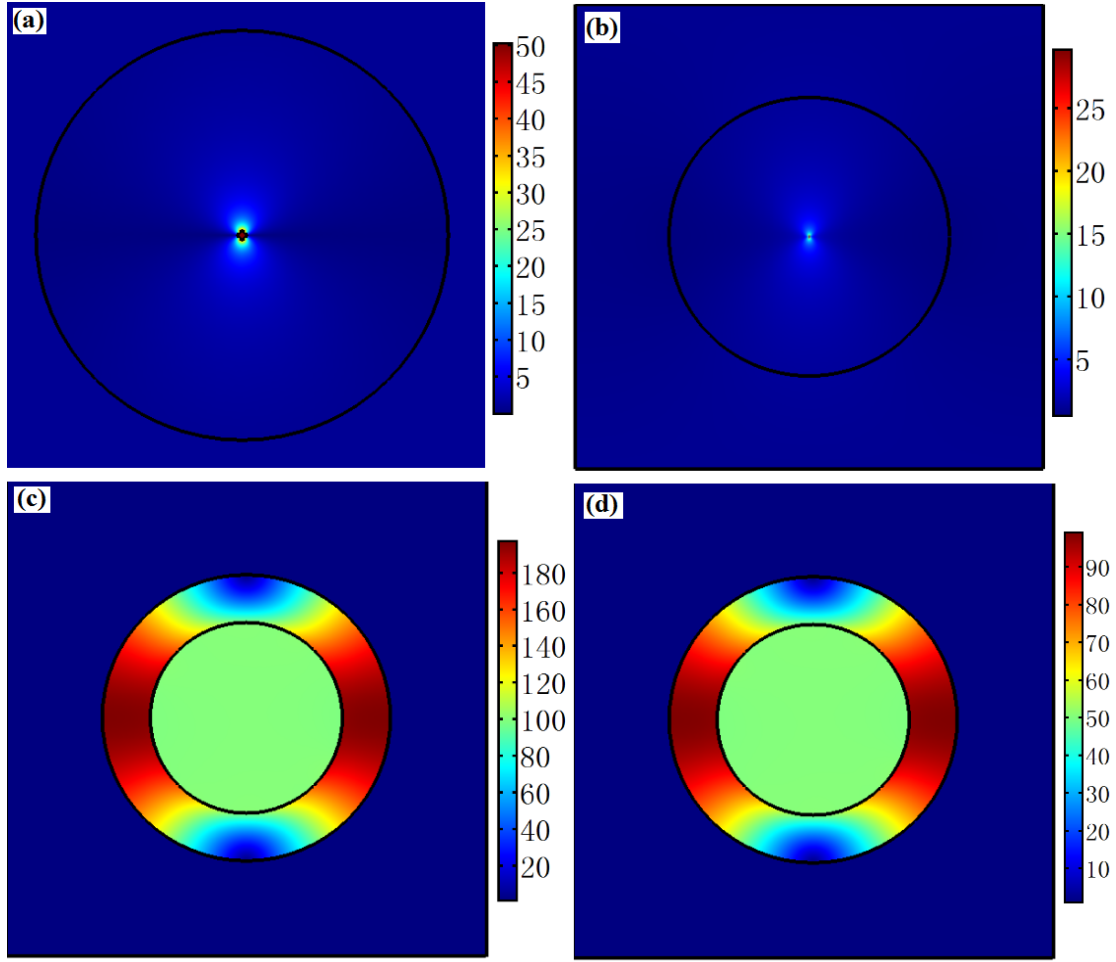

Fig. 2. FEM simulation results. The magnetic flux density distribution in 2D space when we apply a uniform background magnetic flux density $B=1T$ for different devices: (a) the magnetic energy harvesting device proposed in [20]. As we can see here if we want to realize a 50T field in a small air region $R_1=0.1m$, we have to use a very large device whose outer radius is $R_2=1m$ and inner radius is $R_1=0.1m$. (b) a magnetic black hole that can get a high magnetic field at its center (it is not free space but some extremely high permeability material). The radius of this inhomogeneous magnetic device is 0.3m. (c) and (d) our proposed super-concentrator with $R_1=0.2m$ and $R_2=0.3m$. We have a uniform magnetic field of 100T and 50T inside the device (it is in free space region) for (c) $R_3=20m$ and (d) $R_3=10m$, respectively. Note that $R_3$ is a virtual size which affects the degree of enhancement and material distribution of the device. The geometrical size of the device in (c) and (d) is completely determined by $R_1$ and $R_2$.

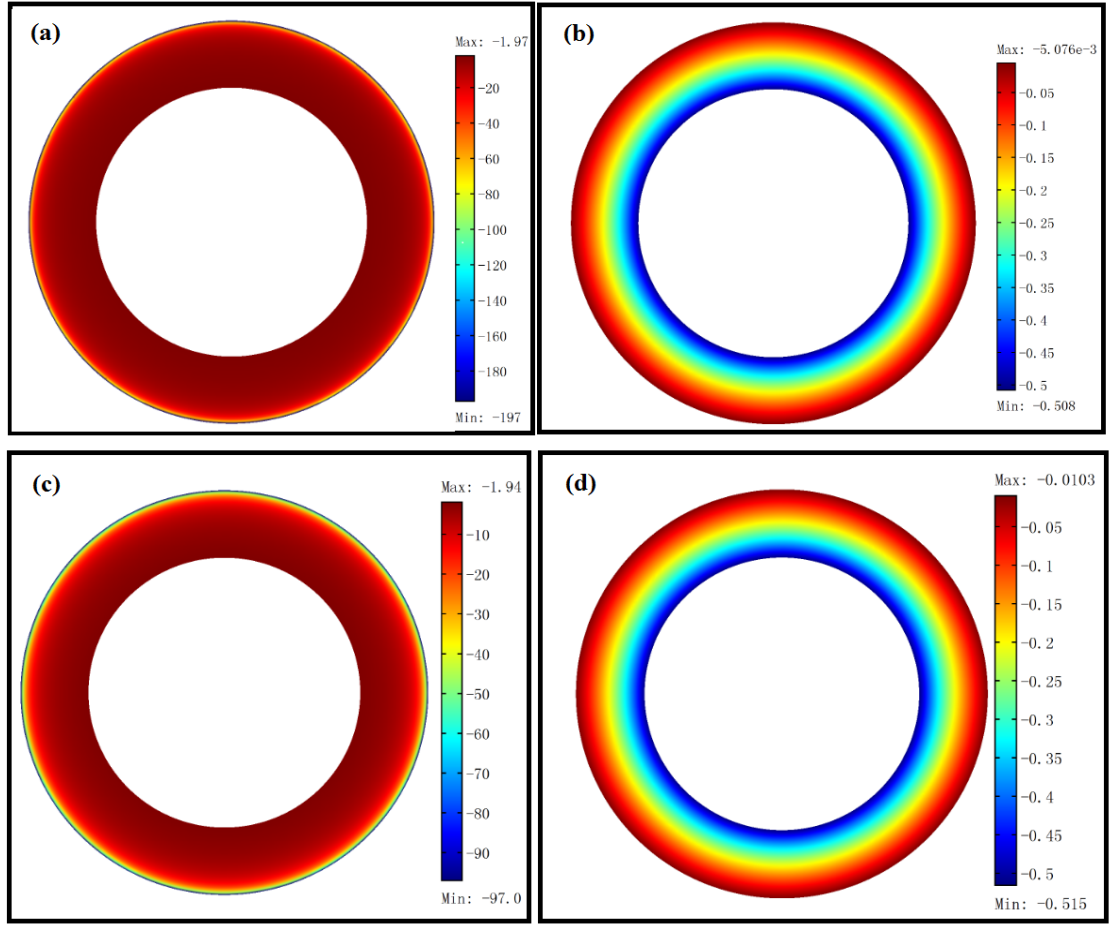

Fig. 3. The permeability of a super-concentrator for a static magnetic field in tangential direction $\mu_\theta$ [(a) and (c)] and in radial direction $\mu_\rho$ [(b) and (d)]. The materials in (a) and (b) are for the proposed device in Fig. 2(c) with $R_1$=0.2m, $R_2$=0.3m and $R_3$=20m. The materials in (c) and (d) are for the proposed device in Fig. 2(d) with $R_1$=0.2m, $R_2$=0.3m and $R_3$=10m.

4. Summary

A compact passive device that can be used to realize a uniform static magnetic field larger than 50T in a large 2D free space has been proposed and designed by transformation optics and verified by the finite element method. This device should have potential applications in many areas where a high magnetic field is required (e.g., magnetic resonance imaging, etc.) over a large free space area. Compared with recent methods for enhancing static magnetic fields, our device is passive and compact, and is made of negative anisotropic inhomogeneous permeability, which can be realized by magnetic metamaterials.

Acknowledgment
This work is partially supported by the National High Technology Research and Development Program (863 Program) of China (No. 2012AA030402), the National Natural Science Foundation of China (Nos. 61178062 and 60990322), the Program of Zhejiang Leading Team of Science and Technology Innovation, Swedish VR grant (# 621-2011-4620) and AOARD.